\documentclass[aps,pra,reprint,superscriptaddress,twocolumn]{revtex4}
\usepackage[dvipdfmx]{graphicx}
\usepackage{graphicx}
\usepackage{color}
\usepackage[T1]{fontenc}
\usepackage{textcomp}
\usepackage{bm}
\usepackage{amsmath,amssymb,amsthm}
\usepackage{comment}
\usepackage{ulem}
\usepackage{grffile}
\usepackage{algorithm}
\usepackage{algpseudocode}

\newcommand{\cket}[1]{\left|#1\right\rangle}
\newcommand{\bra}[1]{\left\langle#1\right|}

\begin{document}


\title{Nonergodic dynamics of the one-dimensional Bose-Hubbard model with a trapping potential}

\author{Masaya Kunimi}
\email{kunimi@ims.ac.jp}
\affiliation{Department of Photo-Molecular Science, Institute for Molecular Science, National Institutes of Natural Sciences, Myodaiji, Okazaki 444-8585, Japan}
\author{Ippei Danshita}
\email{danshita@phys.kindai.ac.jp}
\affiliation{Department of Physics, Kindai University, Higashi-Osaka, Osaka 577-8502, Japan}


\date{\today}

\begin{abstract}
We investigate nonergodic behavior of the one-dimensional Bose-Hubbard model, which emerges in the unitary quantum dynamics starting with initial-state $|\psi(0)\rangle=|\cdots 2020\cdots \rangle$ in the presence of a trapping potential. We compute the level spacing statistic, the time evolution of the number imbalance between the odd and the even sites, and the entanglement entropy in order to show that the system exhibits nonergodicity in a strongly interacting regime. The trapping potential enhances nonergodicity even when the trapping potential is weak compared to the the hopping energy. We derive the effective spin-1/2 {\it XXZ} Hamiltonian for the strongly interacting regimes by using a perturbation method. On the basis of the effective Hamiltonian, we show that the trapping potential is effectively strengthened by the on-site interaction, leading to the enhancement of the nonergodic behavior. We also calculate the real-time dynamics under the effective Hamiltonian and find that the entanglement entropy grows logarithmically in time.
\end{abstract}

\maketitle

\section{Introduction}\label{sec:Introduction}

Thanks to the development of artificial quantum systems, such as ultracold gases, trapped ions, and Rydberg atoms, the problem of thermalization of isolated quantum systems has attracted much attention in the past decade. The key concept of this problem is the eigenstate thermalization hypothesis (ETH) \cite{Deutsch1991,Srednicki1994,Rigol2008}, which states that the expectation value of a few-body operator with respect to an eigenstate of the Hamiltonian of a quantum many-body system coincides with that of the microcanonical ensemble. When the system satisfies the strong version of the ETH (all eigenstates satisfy the ETH), the system thermalizes after long-time unitary evolution \cite{D'Alessio2016,Mori2018}. Thus, the ETH can be regarded as the quantum version of the ergodicity hypothesis. 

Integrable systems \cite{Rigol2007,Rigol2009,Rigol2009_2,Cassidy2011,Vidmar2016} and many-body localized systems \cite{Gorniyi2005,Basko2006,Huse2014,Nandkishore2015,Altman2015,Sierant2017,Sierant2017_2,Sierant2018,Abanin2019} are known as systems that do not satisfy the ETH. The integrable systems do not thermalize due to the presence of an extensive number of conserved quantities. The thermal equilibrium states of the integrable systems can be described by the generalized Gibbs ensemble \cite{Rigol2007,Cassidy2011,Vidmar2016}. The many-body localization (MBL) phase appears in the presence of a strongly disordered potential. Although the MBL systems are not integrable, there is an extensive number of conserved quantities due to the strong disorder. Therefore, the MBL systems do not thermalize. 

Recently, novel types of nonergodic systems in the absence of disordered potentials have been found theoretically and experimentally, such as quantum many-body scar \cite{Turner2018,Turner2018_2,Serbyn2021}, Stark MBL \cite{Schulz2019,Nieuwenburg2019}, and Hilbert space fragmentation (or shattering) \cite{Moudgalya2019_a,Sala2020,Khemani2020}. In systems where the quantum many-body scar can occur, some eigenstates (scar states) that violate the ETH exist. If we choose a state overlapped largely with the scar states, the system does not thermalize. This phenomenon has been observed in quantum simulators using Rydberg atoms in an optical tweezer array \cite{Bernien2017,Bluvstein2021}. The Stark MBL occurs in the presence of a strong linear potential and an additional weak nonlinear potential. The properties of the Stark MBL are similar to those of the disordered MBL. The stark MBL has been observed in trapped-ion systems \cite{Morong2021a}. The Hilbert space fragmentation occurs when some dynamical constraints are present. The Hilbert space is then fragmented into an exponentially large number of subsectors, which are not related to symmetries of the system. Because the dynamics are restricted in one or a few of these subsectors, the system does not thermalize. This phenomenon has been observed in a one-dimensional (1D) Fermi gas in an optical lattice, which serves as a quantum simulator of the Fermi-Hubbard model, in the presence of a strong linear potential~\cite{Scherg2020a,Kohlert2021a}. 

In this paper, we study nonergodic behavior of the 1D  Bose-Hubbard model. The previous works investigated level-spacing statistics, expectation values of physical quantities of the eigenstates and dynamics, and showed that strong interparticle interactions lead to the nonergodicity even when the system is homogeneous \cite{Kollath2010,Roux2010,Carleo2012,Sorg2014,Russomanno2020}. From these works, it can be expected that the nonergodic behavior is likely due to the Hilbert space fragmentation. From an experimental point of view, it is important to consider the effects of a trapping potential, and there are several previous works along this direction \cite{Chanda2020,Yao2020,Yao2021a}. In particular, Yao and co-workers considered the dynamics of the 1D Bose-Hubbard model with a strong parabolic potential with initial-state $\cket{\psi(0)}=\cket{\cdots 1010\cdots}$ \cite{Yao2020,Yao2021a}. They showed that the system exhibits the MBL-like behavior when the strength of the trapping potential is strong enough. They also found that the growth of the entanglement entropy (EE) is logarithmic in time.

Unlike the previous work by Yao and co-workers, we focus on the initial condition $\cket{\psi(0)}=\cket{\cdots 2020\cdots}$. In this case, nonergodic dynamics have been found in a strongly interacting regime when there is no trapping potential~\cite{Carleo2012, Russomanno2020}. We show that a weak trapping potential enhances the nonergodic behavior caused by the large interaction. To understand this property, we derive the effective Hamiltonian by using a perturbation method, which is valid for the strongly interacting regime. The effective Hamiltonian becomes the spin-1/2 {\it XXZ} model with nonuniform spin exchange and a longitudinal magnetic field. Based on the effective Hamiltonian, we discuss the origin of the nonergodic behavior in the case that the trapping potential is weak. We also find the logarithmic growth of the EE in time for some parameters. 

This paper is organized as follows: In Sec.~\ref{sec:model}, we explain our model. In Sec.~\ref{subsec:small_system}, we show the results of small systems based on calculations using the full Hilbert space, which we call the exact diagonalization (ED). The level-spacing statistics and the real-time evolution are discussed. In Sec.~\ref{subsec:large_system}, we show the results of relatively large systems based on the matrix product state (MPS) calculations. We calculate the real-time dynamics of the imbalance and EE. We also derive the effective {\it XXZ} Hamiltonian and show the results of real-time dynamics under the effective Hamiltonian. In Sec.~\ref{sec:summary}, we summarize our results. In the Appendix, we explain the details of the Schrieffer-Wolff transformation. 

\section{Model}\label{sec:model}

We consider the Bose-Hubbard model~\cite{Fisher1989} on a 1D open chain. This model quantitatively describes ultracold 1D Bose gases in a sufficiently deep optical lattice~\cite{Jaksch1998, Greiner2002}. The Hamiltonian is given by
\begin{align}
\hat{H}=-J\sum_{\langle i,j\rangle}\hat{a}^{\dagger}_{i}\hat{a}_j+\sum_{i=1}^MV_i\hat{n}_i+\frac{U}{2}\sum_{i=1}^M\hat{n}_i(\hat{n}_i-1),\label{eq:Hamiltonian}
\end{align}
where $\hat{a}_i (\hat{a}_i^{\dagger})$ is an annihilation (creation) operator of a boson at site $i$, $\hat{n}_i\equiv \hat{a}_i^{\dagger}\hat{a}_i$ is a number operator at site $i$, $J>0$ is a hopping amplitude, $U>0$ is an on-site interaction strength, $\langle i, j\rangle $ represents the summation over nearest-neighbor sites, $M$ is the number of lattice sites, $V_i=\Omega[i-(M+1)/2]^2$ is the parabolic potential, and $\Omega>0$ is the strength of the parabolic potential. For simplicity, we consider the even-$M$ case only. 

We use a number basis $\cket{\bm{n}}\equiv \cket{n_1,n_2,\ldots n_M}$, where $n_i=0,1,\ldots$ is the number of atoms at site $i$. Because the Hamiltonian (\ref{eq:Hamiltonian}) has $U(1)$ symmetry, we restrict the basis $\cket{\bm{n}}$ in the Hilbert subspace $\mathcal{H}_N\equiv \{\cket{\phi} | \hat{N}\cket{\phi}=N\cket{\phi}\}$ where $\hat{N}\equiv \sum_i\hat{n}_i$ is the total number operator and $N$ is the total number of atoms in the system. In addition to the $U(1)$ symmetry, the system has the space-inversion symmetry. The Bose-Hubbard Hamiltonian (\ref{eq:Hamiltonian}) commutes with the space inversion operator $\hat{\mathcal{I}}$, which is defined by $\hat{\mathcal{I}}\hat{a}_i\hat{\mathcal{I}}^{-1}=\hat{a}_{M-i+1}$. Throughout this paper, we use the $U(1)$ symmetry to reduce the computational cost. In Sec.~\ref{subsec:small_system}, we use both $U(1)$ and space inversion symmetry to diagonalize the Hamiltonian (\ref{eq:Hamiltonian}) in the Hilbert subspace $\mathcal{H}_{N,\mathcal{I}}=\{\cket{\phi} | \; \hat{N}\cket{\phi}=N\cket{\phi},\;\hat{\mathcal{I}}\cket{\phi}=\mathcal{I}\cket{\phi}\}$, where $\mathcal{I}=\pm1$ is the eigenvalue of the space inversion operator. See the details of the ED in Refs.~\cite{Sandvik2010,Jung2020}.

\section{Results}\label{sec:results}

\subsection{Small systems}\label{subsec:small_system}

In this section, we present the results obtained by ED for $M=N=10$. First, we show the level spacing statistics. We calculate the mean level spacing ratio $r$ defined as \cite{Oganesyan2007,Atas2013}
\begin{align}
r\equiv \left\langle\min(r_n,1/r_n)\right\rangle,\label{eq:definition_of_mean_level_spacing_latio}
\end{align}
where $r_n\equiv s_{n+1}/s_n$, $s_n\equiv E_{n+1}-E_n$, $E_n$ is the $n$th eigenenergy of the Hamiltonian in the Hilbert subspace $\mathcal{H}_{N,\mathcal{I}}$ and $\langle\cdots\rangle$ represents the average over all states, respectively. Here, we sort the eigenenergy in the ascending order: $E_1\le E_2\le\cdots$. The mean level spacing ratio is an indicator of the ergodicity. If the system is ergodic, the system is well described by the random matrix theory. Because the Bose-Hubbard model is spinless and has time-reversal symmetry, we expect that the system has similar features to the Gaussian orthogonal ensemble. In this case, the mean level spacing ratio becomes $r\simeq 0.53$. In contrast to this, if the system is localized or integrable, the level spacing statistics obey the Poisson distribution. In this case, the mean level spacing ratio is given by $r\simeq 0.38$.

\begin{figure}[t]
\centering
\includegraphics[width=8.5cm,clip]{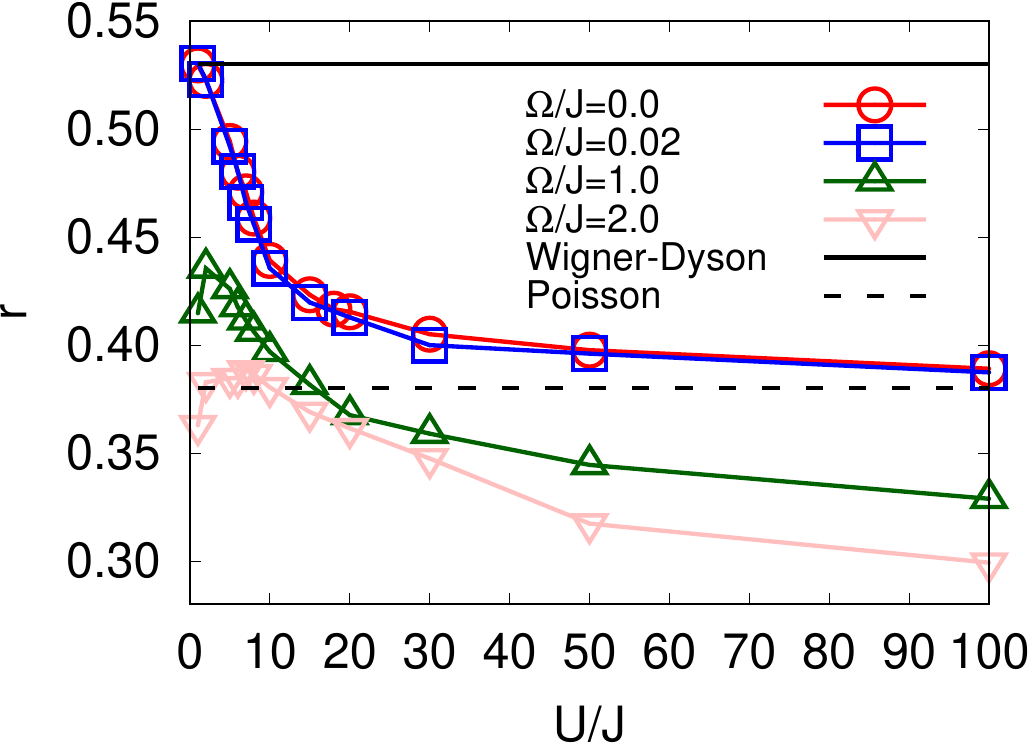}
\caption{Mean level spacing ratio for the Bose-Hubbard model of $M=N=10$ and $\mathcal{I}=+1$ as a function of $U/J$. The black solid dashed and dotted lines represent the predicted values of the Wigner-Dyson and Poisson distributions. In the case of $\mathcal{I}=-1$, the results are almost the same as $\mathcal{I}=+1$ (data are not shown).}
\label{fig:level-spacing_statistic}
\end{figure}%

Figure \ref{fig:level-spacing_statistic} shows the mean level spacing ratio versus $U/J$ for several values of $\Omega/J$ in the Hilbert subspace $\mathcal{H}_{N,\mathcal{I}}$ with $N=10$ and $\mathcal{I}=+1$. We see that when the trapping potential is as weak as $\Omega/J\leq 0.02$, $r$ is close to the Wigner-Dyson (Poisson) value in a region of small (large) $U/J$. This behavior is consistent with the previous work in the absence of the trapping potential \cite{Kollath2010,Russomanno2020}. As we will see below, the nonergodic behavior appears in the Poisson distribution regime. This behavior can be understood by the Hilbert space fragmentation induced by the large on-site interaction. To see this, we consider the atomic limit ($J=\Omega=0$). In this case, the energy eigenstate of the Hamiltonian (\ref{eq:Hamiltonian}) is given by $\cket{\bm{n}}$ and can be characterized by the partition of $N$, which is a way of expressing $N$ as a sum of positive integers. The particle number $N$ can be written by $N=\sum_{j=0}^Njl_j$, where $l_j$ is the number of $n_i$ in $\cket{\bm{n}}$ that satisfies $n_i=j$. When the different states $\cket{\bm{n}}$ and $\cket{\bm{n}'}$ have the same $\bm{l}\equiv (l_0,l_1,\ldots, l_N)$, the eigenenergy of these states is the same. This means that we can label the eigenstates by $\bm{l}$. The asymptotic form of the total number of the partition is given by $p(N)\sim [1/(4N\sqrt{3})]\exp{(\pi\sqrt{2N/3})}]$ \cite{Hardy1918}. This means that an exponentially large number of sectors exist. In the case that $0<J\ll U$ and $0<\Omega\ll U$, the different sectors are coupled due to the hopping term. Nevertheless, the above structure will survive because there is a large energy gap between the different subsectors as long as $U/J$ is large. Therefore, the nonergodicity in the Poisson distribution regime for small $\Omega/J$ can be attributed to the Hilbert space fragmentation induced by the large on-site interaction. On the other hand, when $\Omega/J\geq 1.0$, $r$ significantly deviates from the Wigner-Dyson value for all values of $U/J$. It also deviates from the Poisson value for large values of $U/J$. One possible reason for this behavior is the emergence of the quadrupole conservation law, which leads to the Hilbert space fragmentation induced by large $\Omega$ \cite{Khemani2020,Sala2020,Feldmeier2020,Gromov2020}. 

\begin{figure}[t]
\centering
\includegraphics[width=8.5cm,clip]{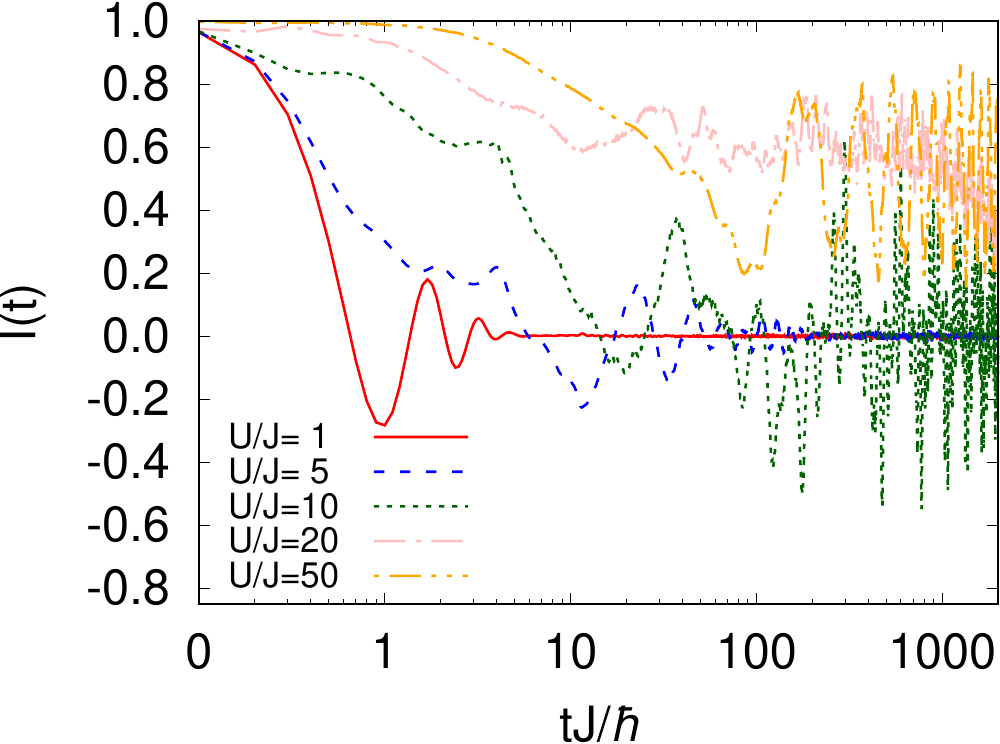}
\caption{Time evolution of the imbalance of the Bose-Hubbard model of $M=N=10$ and $\Omega=0.02J$.}
\label{fig:imbalance_small_size}
\end{figure}%

\begin{figure}[t]
\centering
\includegraphics[width=8.5cm,clip]{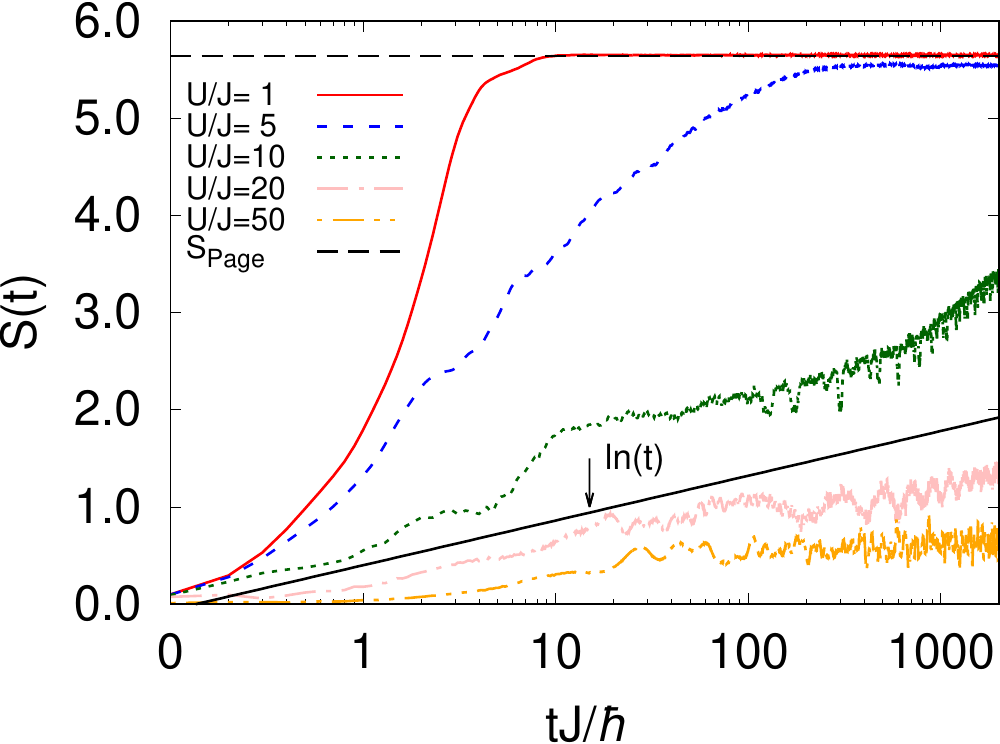}
\caption{Time evolution of the entanglement entropy of the Bose-Hubbard model of $M=N=10$ and $\Omega=0.02J$.  The black dashed line represents the Page value. The black solid line represents $\ln (t)$ for a guide to the eye.}
\label{fig:entanglement_small_size}
\end{figure}%

Next, we calculate the time evolution by solving the time-dependent Schr\"odinger equation 
\begin{align}
i\hbar\frac{d}{d t}\cket{\psi(t)}&=\hat{H}\cket{\psi(t)},\label{eq:time-dependent_schrodinger_equation}
\end{align}
where $\cket{\psi(t)}$ is a state vector at time $t$. The initial condition is $\cket{\psi(0)}=\cket{2020\cdots}$. We use the fourth-order Runge-Kutta method. Here, we focus on two physical quantities. One is a particle number imbalance $I(t)\equiv [N_{\rm odd}(t)-N_{\rm even}(t)]/N$, where $N_{\rm odd}(t)\equiv \bra{\psi(t)}\sum_{i=1,3\cdots}\hat{n}_i\cket{\psi(t)}$ and $N_{\rm even}(t)\equiv \bra{\psi(t)}\sum_{i=2,4\cdots}\hat{n}_i\cket{\psi(t)}$ are the particle numbers of odd and even sites, respectively. This quantity is a good indicator of the nonergodicity because the imbalance becomes zero in the thermal equilibrium states. Another advantage of the imbalance is that it can be experimentally measured by using the band mapping technique with optical superlattices \cite{Scherg2020a,Kohlert2021a,Trotzky2012}. The other quantity is the von Neumann EE: $S(t)\equiv -{\rm Tr}\{\hat{\rho}_A(t)\ln[\hat{\rho}_A(t)]\}$, where $\hat{\rho}_A(t)\equiv {\rm Tr}_B[\hat{\rho}(t)]$ is the reduced density matrix of the subsystem $A$, ${\rm Tr}_B$ denotes the trace over the subsystem $B$ (complement of the subsystem $A$), and $\hat{\rho}(t)\equiv \cket{\psi(t)}\bra{\psi(t)}$ is the density matrix of the whole system. Throughout this paper, we consider a bipartite EE, that is, the subsystem $A$ is always taken to be the half-left of the system.

Figures \ref{fig:imbalance_small_size} and \ref{fig:entanglement_small_size} show the time evolution of the imbalance and EE for $\Omega=0.02J$. We can see that the imbalance decays to zero for small values of $U/J$, which indicates that thermalization occurs. The thermalization can also be seen in the EE. The EE for $U/J=1$ and $5$ almost saturates at the Page value $S_{\rm Page}$, which is an EE of the random state vector \cite{Page1993}. In the present case, we calculate the Page value by generating $1000$ random normalized vectors in the Hilbert space $\mathcal{H}_N$. We omit the error bars of the Page value in Fig.~\ref{fig:entanglement_small_size} because the statistical uncertainty is sufficiently small.

On the contrary to the case that $U/J$ is small, the nonergodic behavior appears when $U/J$ is large. In Fig.~\ref{fig:imbalance_small_size}, we can see that when $U/J\ge 20$, the imbalance noticeably deviates from zero, at least, up to $t=2000\hbar/J$. As seen in Fig.~\ref{fig:entanglement_small_size}, the nonergodicity of the dynamics is evident also in the EE. The growth of the EE is quite slow. In particular, we can see the logarithmic growth of the EE for $U/J=20$. This logarithmic growth of the EE is similar to the disordered MBL cases \cite{Abanin2019}. We note that this logarithmic growth of the EE under the disorder-free potential has been found in Refs. \cite{Chanda2020,Yao2020,Yao2021a}.

\subsection{Large systems}\label{subsec:large_system}

In this section, we present the results obtained by MPS for relatively large systems. We take $M=50$ and $\Omega=0.02J$. These values are evaluated by a realistic experimental setup \cite{Takasu2020}. The strength of the trapping potential $\Omega=0.02J$ roughly corresponds to a trap frequency $\omega/2\pi\sim 30$ Hz. To perform the calculations of real-time dynamics in large systems by means of MPS, we use the time-evolving block decimation (TEBD) algorithm \cite{Vidal2003,Vidal2004} exploiting $U(1)$ symmetry \cite{Singh2010}. We restrict the dimension of the local Hilbert space $d_{\rm max}=5$-$9$ (the maximum occupation number is $n_{\rm max}=d_{\rm max}-1$) and use the maximum bond dimension $\chi_{\rm max}=100$-$1600$, which depend on the parameters.
The initial condition for the real-time evolution is{$\cket{\psi(0)}=\cket{\cdots2020\cdots}$. Here, the particles are placed only near the central part of the system in order to mimic the actual experiments.

\begin{figure}[t]
\centering
\includegraphics[width=8.5cm,clip]{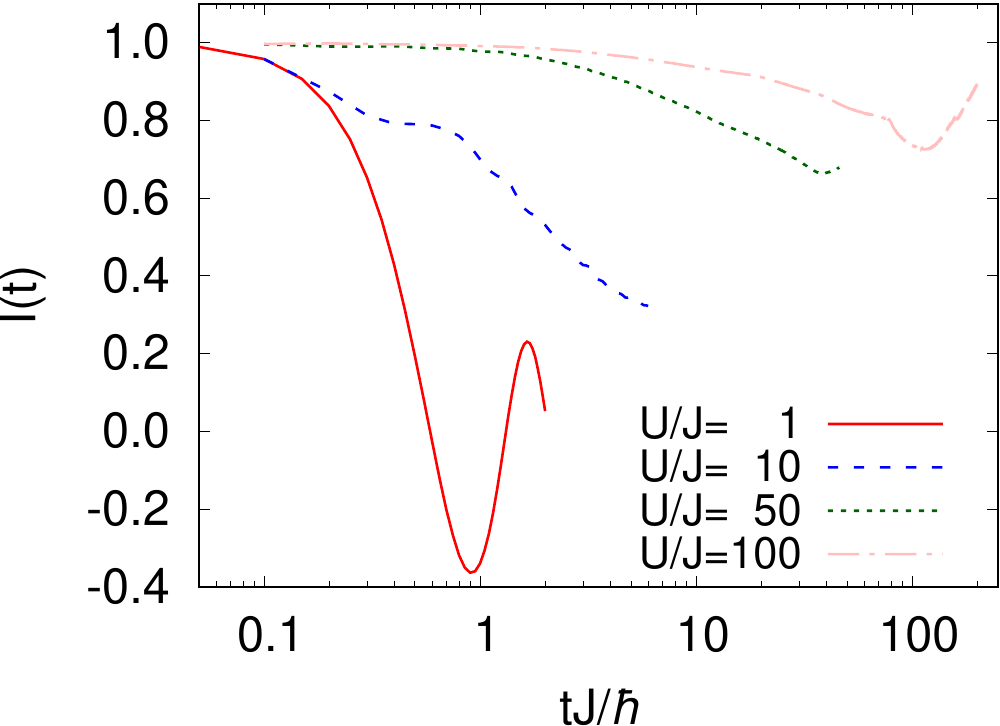}
\caption{Time evolution of the imbalance of the Bose-Hubbard model for $M=50$, $N=32$, and $\Omega=0.02J$.}
\label{fig:imbalance_02_large_size}
\end{figure}%

\begin{figure}[t]
\centering
\includegraphics[width=8.5cm,clip]{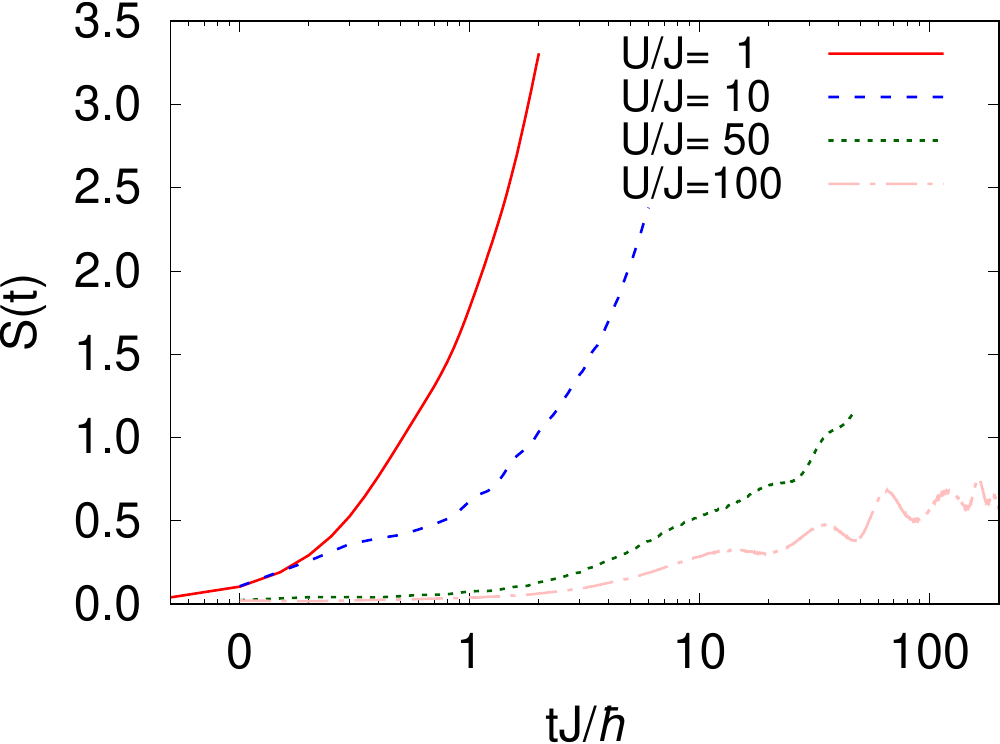}
\caption{Time evolution of the entanglement entropy of the Bose-Hubbard model for $M=50$, $N=32$, and $\Omega=0.02J$.}
\label{fig:entanglement_large_size}
\end{figure}%

Figure \ref{fig:imbalance_02_large_size} shows the time evolution of the imbalance for $M=50$, $N=32$, and $\Omega=0.02J$. We can see that the slow decay of the imbalance for the large values of $U/J$. We can also find that the growth of the EE for the large $U/J$ is quite slow (see Fig.~\ref{fig:entanglement_large_size}). These results suggest that the system is nonergodic for large $U/J$. On the other hand, we can see the fast decay of the imbalance and rapid growth of the EE when $U/J$ is small. This behavior is consistent with the thermalization. However, our calculations for small $U/J$ are limited to short timescale $t\lesssim 10\hbar/J$ because the growth of the EE is fast. We cannot access the fully thermalized states in the small $U/J$ cases.

\begin{figure}[t]
\centering
\includegraphics[width=8.5cm,clip]{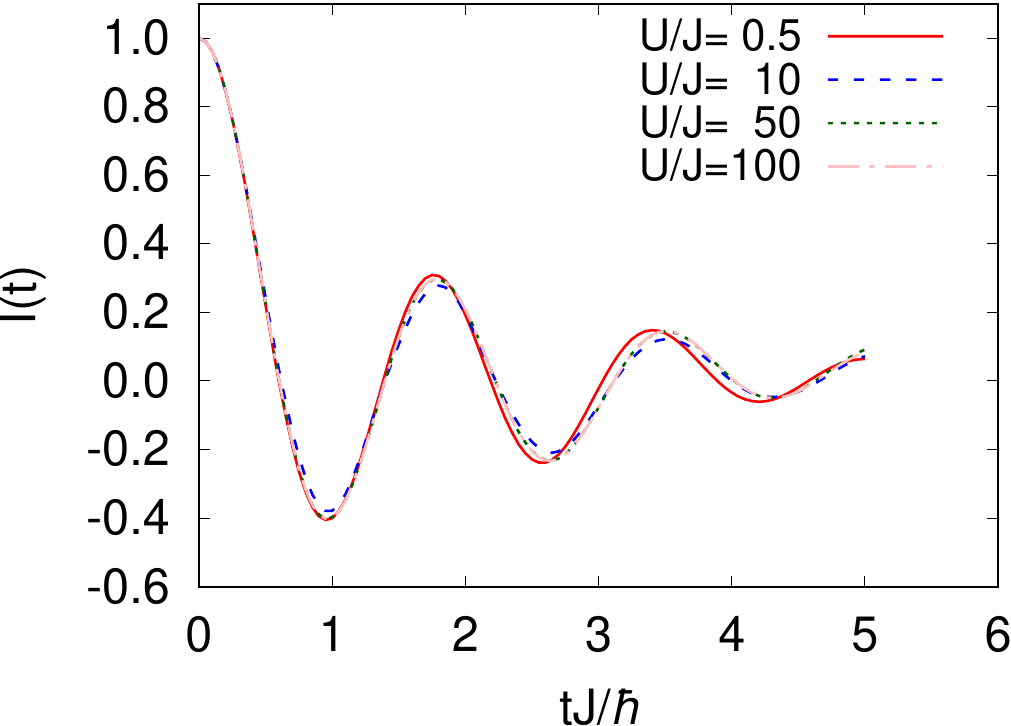}
\caption{Time evolution of the imbalance of the Bose-Hubbard model for $M=50$, $N=16$, and $\Omega=0.02J$.}
\label{fig:imbalance_01_large_size}
\end{figure}%

For comparison, we consider the case that the initial state is $\cket{\psi(0)}=\cket{\cdots 1010\cdots}$, where we set $M=50$, $N=16$, and $\Omega=0.02J$. The time evolution of the imbalance is shown in Fig.~\ref{fig:imbalance_01_large_size}. Comparing with the previous situations, we can see that the decay of the imbalance is fast and the results are almost independent of the interaction. This behavior is qualitatively different from the case that the initial state is $\cket{\psi(0)}=\cket{\cdots2020\cdots}$.

In order to understand this difference, we derive the effective model of the Bose-Hubbard Hamiltonian (\ref{eq:Hamiltonian}) for large $U/J$ using the Schrieffer-Wolff transformation \cite{Cohen-Tannoudji_book,Bravyi2011}. Here, we consider the Hilbert subspace $\mathcal{H}_{02}=\{\cket{\bm{n}}|n_i=0 \text{ or }2, N=\sum_in_i\}$. To write down the effective Hamiltonian, we introduce the spin-1/2 operators; $\hat{S}_i^z\equiv (1/2)(\cket{2_i}\bra{2_i}-\cket{0_i}\bra{0_i})$, $\hat{S}_i^+\equiv \cket{2_i}\bra{0_i}=(\hat{S}_i^-)^{\dagger}$, $\hat{S}_i^x\equiv (\hat{S}_i^++\hat{S}_i^-)/2$, and $\hat{S}_i^y\equiv (\hat{S}_i^+-\hat{S}_i^-)/(2i)$. This mapping means that we regard $\cket{2_i}$ as $\cket{\uparrow_i}$ and $\cket{0}_i$ as $\cket{\downarrow_i}$, where $\cket{\uparrow_i}$ and $\cket{\downarrow_i}$ are the eigenstates of $\hat{S}_i^z$. Using these operators, we can write down the effective Hamiltonian in the spin language (see the Appendix),
\begin{align}
\hat{H}_{\rm eff}^{02}&=E_0+\sum_{i=1}^Mh_i^z\hat{S}_i^z\notag \\
&\;+2\sum_{i=1}^{M-1}\tilde{J}_{i,i+1}(\hat{S}_{i+1}^x\hat{S}_i^x+\hat{S}_{i+1}^y\hat{S}_i^y-4\hat{S}_{i+1}^z\hat{S}_i^z),\label{eq:second_order_effective_Hamiotonian_for_trap}\\
\tilde{J}_{i,i+1}&\equiv 
\begin{cases}
\vspace{0.5em}\dfrac{2J^2U}{U^2-(V_{i+1}-V_i)^2},\quad i=1,\cdots, M-1,\\
0,\hspace{8.5em} i=0,\text{ or }M,
\end{cases}
\label{eq:definition_of_exchange_coupling_trap_tilde}\\
E_0&\equiv \frac{1}{2}MU+\sum_{i=1}^MV_i-\sum_{i=1}^{M-1}\tilde{J}_{i,i+1},\label{eq:definition_of_E0}\\
h_i^z&\equiv 2V_i-2(\tilde{J}_{i-1,i}+\tilde{J}_{i,i+1})-2J^+_{i,i+1}-2J^-_{i-1,i},\label{eq:definition_of_hz_for_trap_XXZ}\\
J_{i,i+1}^{\pm}&\equiv
\begin{cases}
\vspace{0.5em}\dfrac{J^2}{U\pm(V_{i+1}-V_i)},\quad i=1,\cdots, M-1,\\
0,\hspace{7.5em}i=0,\text{ or } M.
\end{cases}
\label{eq:definition_exchange_coupling_trap}
\end{align}
This result shows that the effective Hamiltonian is a spin-1/2 {\it XXZ} model with a nonuniform exchange coupling and nonuniform magnetic field along the $z$ direction. We note that the effective Hamiltonian for $\Omega = 0$ has been derived in Refs.~\cite{Carleo2012,Russomanno2020}. Our results can be regarded as the extension of these previous works that include inhomogeneous potentials. We also note that the effective Hamiltonian of two-component bosons in a trapping potential has been derived in Ref.~\cite{Garcia-Ripoll2003}. 

Here, we discuss the validity of the perturbation theory. From Eqs.~(\ref{eq:definition_of_exchange_coupling_trap_tilde}) and (\ref{eq:definition_exchange_coupling_trap}), if $U\simeq|V_{i+1}-V_i|$ is satisfied, the perturbation theory breaks down because state $\cket{1_i1_{i+1}}$ has almost the same energy as the $\cket{0_i2_{i+1}}$ or $\cket{2_i0_{i+1}}$ states. In the following, we consider case $U\gg |V_{i+1}-V_i|$ for all $i$, which we refer to as the weak trap. Note that the similar situations have been discussed in Ref.~\cite{Yao2020}.

In a similar manner, we can derive the effective Hamiltonian in the Hilbert subspace $\mathcal{H}_{01}\equiv \{\cket{\bm{n}}|n_i=0\text{ or }1, N=\sum_in_i\}$. The effective Hamiltonian is given by
\begin{align}
\hat{H}_{\rm eff}^{01}&=E_0'+\sum_{i=1}^Mh_i'^z\hat{\tau}_i^z\notag \\
&-2J\sum_{i=1}^{M-1}(\hat{\tau}_{i+1}^x\hat{\tau}_i^x+\hat{\tau}_{i+1}^y\hat{\tau}_i^y)-2\sum_{i=1}^{M-1}\tilde{J}_{i,i+1}\hat{\tau}_{i+1}^z\hat{\tau}_i^z\label{eq:effective_Hamiltonian_in_H01}\\
E_0'&\equiv \sum_{i=1}^M\left(\frac{V_i}{2}-\frac{\tilde{J}_{i,i+1}}{2}\right),\label{eq:definition_of_E0_in_H01}\\
h'^z_i&=V_i-\tilde{J}_{i-1,i}-\tilde{J}_{i,i+1},\label{eq:definition_of_magnetic_field_in_H01}
\end{align}
where we defined the other spin-1/2 operators by $\hat{\tau}_i^z\equiv (1/2)(\cket{1_i}\bra{1_i}-\cket{0_i}\bra{0_i})$, $\hat{\tau}_i^+\equiv \cket{1_i}\bra{0_i}=(\hat{\tau}_i^-)^{\dagger}$, $\hat{\tau}_i^x\equiv (\hat{\tau}_i^++\hat{\tau}_i^-)/2$, and $\hat{\tau}_i^y\equiv (\hat{\tau}_i^+-\hat{\tau}_i^-)/(2i)$.
 
Here, we discuss the nonergodic behavior from a viewpoint of the effective Hamiltonians (\ref{eq:second_order_effective_Hamiotonian_for_trap}) and (\ref{eq:definition_of_E0_in_H01}). The hopping process and nearest-neighbor density-density interaction in $\mathcal{H}_{02} $ (or {\it XY} and {\it ZZ} interactions in the spin language), namely, $\cket{0_i2_{i+1}}\leftrightarrow\cket{2_i0_{i+1}}$ and $\cket{2_i2_{i+1}}\leftrightarrow\cket{2_i2_{i+1}}$, come from the second-order perturbation. The magnitude of these processes scale as $O(J^2/U)$. This means that all the interaction terms vanish in the limit $U\to\infty$. The nonvanishing term is the magnetic-field term, which does not change the spin configuration. We also find that the ratio of the strength of the trapping potential $\Omega$ and the effective hopping is given by $\Omega/(J^2/U)$. This means that the strength of the trapping potential is enhanced by the on-site interaction. In fact, for $\Omega=0.02J$ and $U/J=100$, we obtain $\Omega/(J^2/U)=2$, which means that the strength of the trapping potential and the effective hopping energy are comparable. This large effective trapping potential leads to the enhancement of the nonergodicity. For example, the oscillation of the imbalance across zero that appears in the absence of the trapping potential \cite{Russomanno2020} disappears in the present system due to the presence of the trapping potential, meaning that the particles are localized more significantly. On the other hand, the hopping process in $\mathcal{H}_{01}$, namely, $\cket{0_i1_{i+1}}\leftrightarrow\cket{1_i0_{i+1}}$, comes from the first order perturbation. The ratio of the trapping potential and the hopping energy is $\Omega/J$, which is independent of the interaction $U$. In addition to this fact, the nearest-neighbor interaction between bosons (or the {\it ZZ} interaction) vanishes in the limit $U\to \infty$. Therefore, we do not see the nonergodic behavior of the imbalance in the $N=16$ and $\Omega=0.02J$. Note that the nonergodicity associated with the integrability at $U\rightarrow \infty$ cannot be identified by the imbalance.

In the calculations for the Bose-Hubbard model, the maximum time is $O(100\hbar/J)$. This limitation comes from the large dimension of the local Hilbert space. In order to compute dynamics in a longer timescale, we investigate the effective Hamiltonian (\ref{eq:second_order_effective_Hamiotonian_for_trap}) whose dimension of the local Hilbert space is two. Moreover, in the effective Hamiltonian, the timescale of the effective hopping is much larger than that of the bare hopping. Combining the small dimension of the local Hilbert space and the large timescale of the effective hopping, we can perform the TEBD simulations for a much longer timescale by using the effective Hamiltonian.

\begin{figure}[t]
\centering
\includegraphics[width=8.5cm,clip]{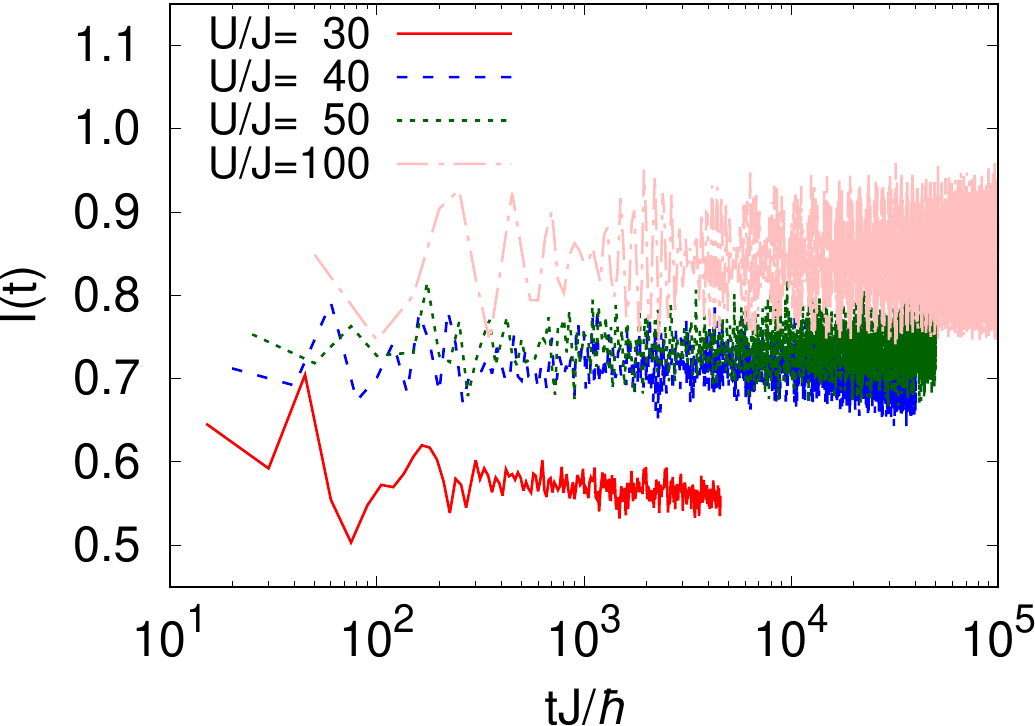}
\caption{Time evolution of the imbalance of the effective {\it XXZ} model (\ref{eq:second_order_effective_Hamiotonian_for_trap}) for $M=50$, $N=32$, and $\Omega=0.02J$.}
\label{fig:imbalance_xxz}
\end{figure}%

\begin{figure}[t]
\centering
\includegraphics[width=8.5cm,clip]{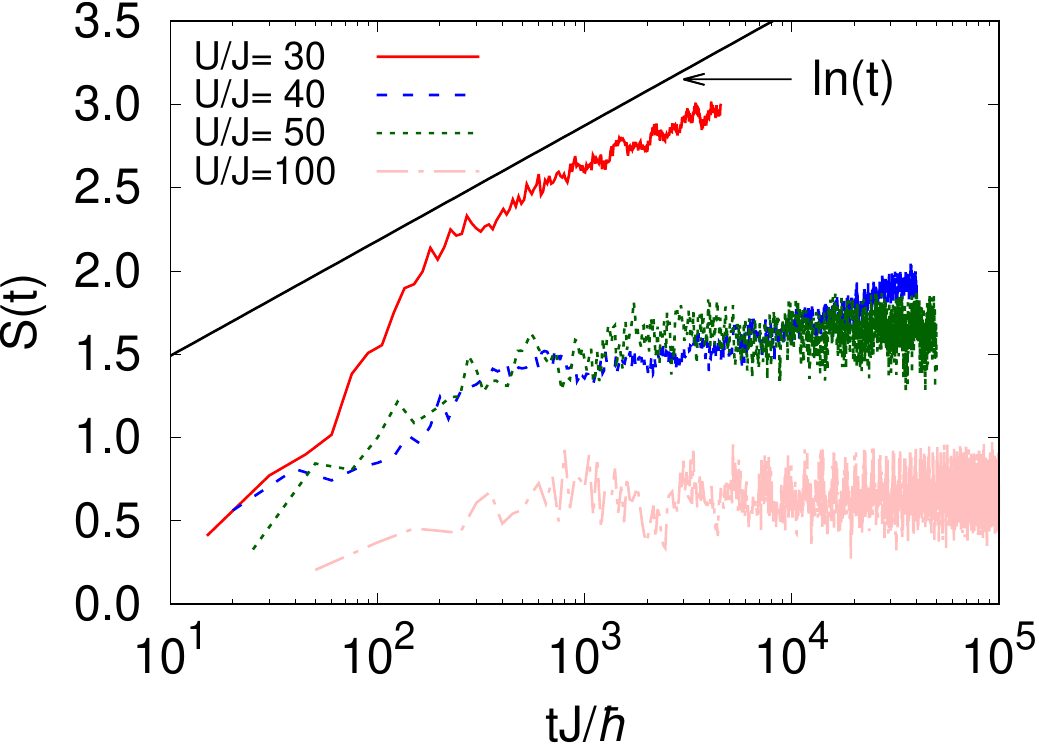}
\caption{Time evolution of the entanglement entropy of the effective {\it XXZ} model (\ref{eq:second_order_effective_Hamiotonian_for_trap}) for $M=50$, $N=32$, and $\Omega=0.02J$. The black solid line represents $\ln (t)$ for a guide to the eye.}
\label{fig:entanglement_xxz}
\end{figure}%

\begin{figure*}[t]
\centering
\includegraphics[width=17cm,clip]{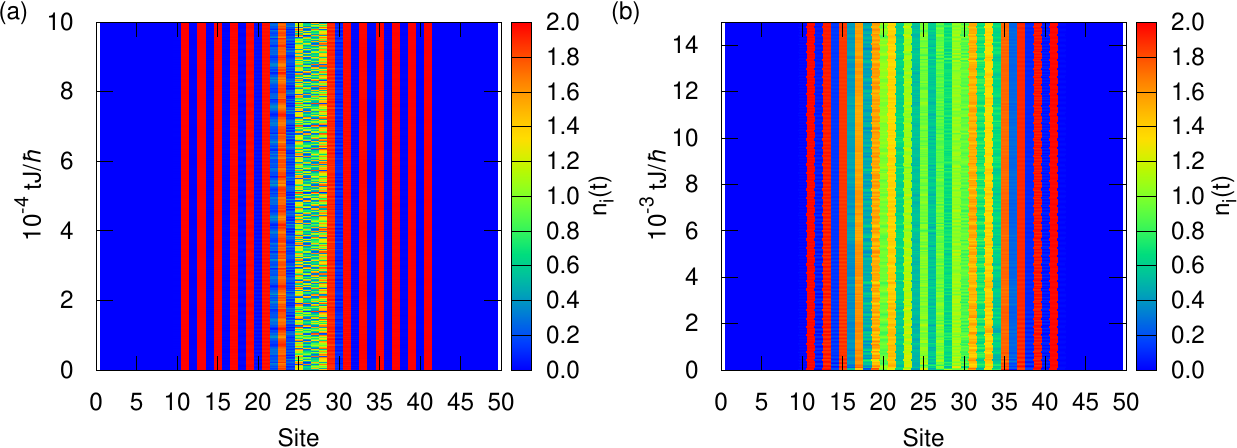}
\caption{Time evolution of the local density [$n_i(t)\equiv \bra{\psi(t)}\hat{n}_i\cket{\psi(t)}=1+2\bra{\psi(t)}\hat{S}_i^z\cket{\psi(t)}$] of the effective {\it XXZ} model for $M=50$, $N=32$, and $\Omega=0.02J$. (a) $U/J=100$ and (b) $U/J=30$.}
\label{fig:density_xxz}
\end{figure*}%

Figure \ref{fig:imbalance_xxz} shows the time evolution of the imbalance of the effective {\it XXZ} model for $M=50$, $N=32$, and $\Omega=0.02J$. We can see that the imbalance remains nonzero up to the timescale $10^3-10^5\hbar/J$. We also show the time evolution of the EE of the {\it XXZ} model in Fig.~\ref{fig:entanglement_xxz}. We can see the logarithmic growth of the EE for $U/J=30$ and $40$, which is similar to MBL in disordered systems. On the other hand, we can see saturation of the EE for $U/J\gtrsim 50$. To understand this behavior, we show the time evolution of the local density profiles. Figure \ref{fig:density_xxz} (a) shows the results for $U/J=100$. In this case, we can see that only particles in a central few sites move during the time evolution. This means that the dynamics are restricted to a small Hilbert subspace, which is the reason why the EE is saturated at a small value for the large $U/J$ cases. In the case that $U/J=30$, mobile particles exist in the central $10-20$ sites [see Fig.~\ref{fig:density_xxz} (b)], but the system still retains the memory of the initial conditions.

\begin{figure}[t]
\centering
\includegraphics[width=8.5cm,clip]{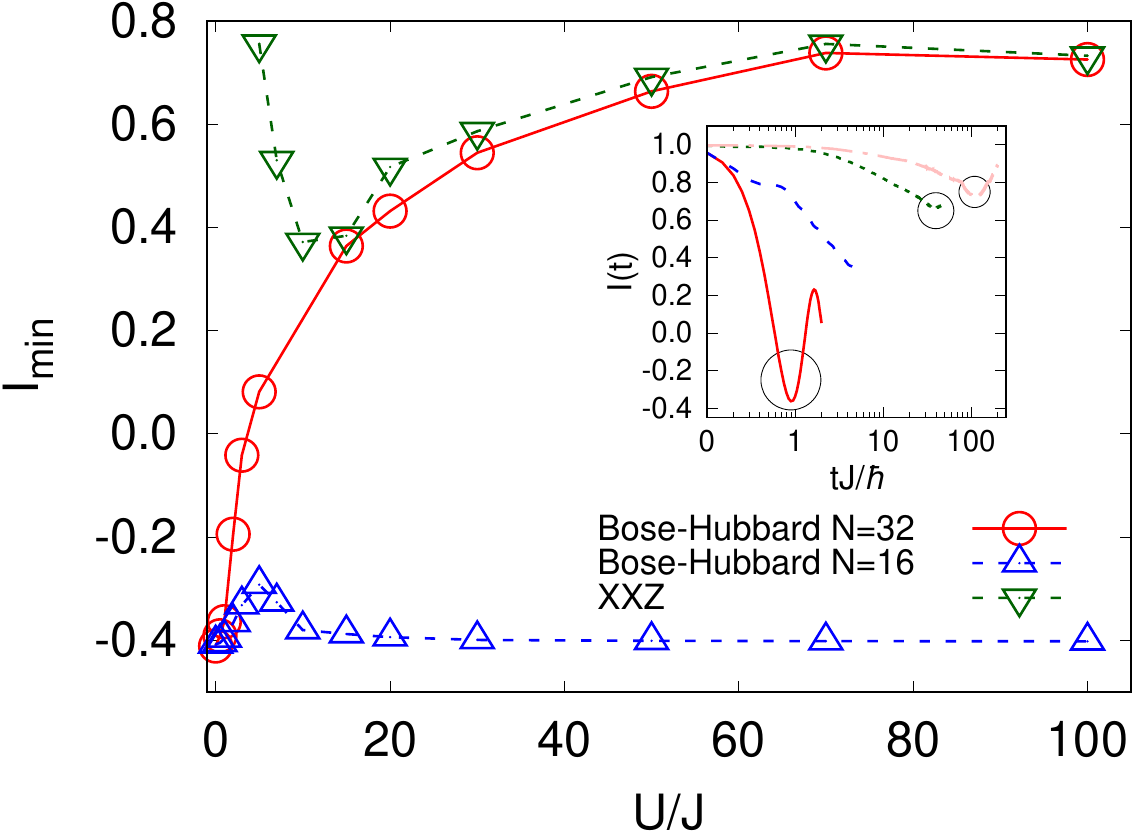}
\caption{$U/J$ dependence of $I_{{\rm min}}$ for $\Omega=0.02J$. The inset shows Fig.~\ref{fig:imbalance_02_large_size}. The black circles in the inset represent $I_{\rm min}$.}
\label{fig:1st_minimum}
\end{figure}%

Here, we discuss the limitations of the timescales of the effective Hamiltonian. According to Ref.~\cite{Carleo2012}, the time evolution under effective Hamiltonian is valid for the timescale shorter than $\hbar(U^2/J^3)$. In the present case, the validity timescales are given by $9\times 10^2-10^4\hbar/J$ for $U/J=30-100$. This means that some of the results are out of these limitations. However, the validity timescales are longer than the maximum time of the Bose-Hubbard calculations [$O(100\hbar/J)$]. In this sense, our calculations are still meaningful. Therefore, we conclude that the nonergodic behavior appears for, at least, $9\times 10^2-10^4\hbar/J$ in the effective {\it XXZ} model.

In order to roughly characterize the nonergodicity from the time evolution of the imbalance, we define $I_{\rm min}$, which is the first minimum value of the imbalance (see the inset in Fig.~\ref{fig:1st_minimum}). Figure \ref{fig:1st_minimum} shows that the $U/J$ dependence of $I_{\rm min}$ of the Bose-Hubbard model for $N=32$ and $16$ and the effective {\it XXZ} model (\ref{eq:second_order_effective_Hamiotonian_for_trap}). We find that $I_{\rm min}$ is negative for all $U/J$'s in $N=16$ Bose-Hubbard models. This indicates that the imbalance will decay to zero after several oscillations. In contrast, the sign of $I_{\rm min}$ changes for $N=32$ Bose-Hubbard models. As shown in this paper, the nonergodic behavior appears in the regime of large $U/J$. We find that $I_{\rm min}$ is close to unity when the system exhibits nonergodicity. From these results, we can roughly capture how nonergodic the dynamics are by the values of $I_{\rm min}$. We also note that we can check the validity of the effective {\it XXZ} model (\ref{eq:second_order_effective_Hamiotonian_for_trap}) from the results of $I_{\rm min}$. We can see that the results of the {\it XXZ} model deviate from those of the $N=32$ Bose-Hubbard model for $U/J\lesssim 20$. This behavior is consistent with the fact that the perturbation theory is valid in the regime of large $U/J$.

Although $I_{\rm min}$ can roughly capture the properties of the nonergodicity, this quantity is insufficient to confirm the nonergodicity. As mentioned in this subsection, the timescales of the numerical simulations of the Bose-Hubbard models are limited due to the computational costs. For a further study, the use of quantum simulators built with ultracold Bose gases in optical lattices will be helpful. Fortunately, the present setup is not difficult to install in the quantum simulator. To perform the quantum simulations of the present systems, we need to prepare initial state $\cket{\cdots 2020\cdots}$ and measure the time dependence of the imbalance. These two operations can be realized by utilizing an optical superlattice technique. For example, we initially prepare the Mott insulator state $\cket{\cdots 1111\cdots}$ in the short optical lattice and then slowly ramp up the long lattice so that the potential minima of sites at odd $j$ are much lower than those at even $j$. After we slowly turn off the hopping by increasing the depth of the short lattice and ramp down the long lattice, we can obtain the $\cket{\cdots 2020\cdots}$ state. The measurements of the imbalance have been performed in some experiments \cite{Trotzky2012,Scherg2020a,Kohlert2021a}. We can naively expect that the nonergodic behavior of the 1D Bose-Hubbard model will be clarified in future experiments. 


\section{Summary}\label{sec:summary}
To summarize, we investigated the nonergodic behavior of the 1D Bose-Hubbard model with a parabolic trapping potential. In small systems, we calculated the level spacing statistics and dynamics starting from state $\cket{\psi(0)}= \cket{2020\cdots}$ by means of the ED. Both results suggest that the nonergodicity appears in the large $U/J$. In relatively large systems, we analyzed the real-time dynamics of the Bose-Hubbard models starting from the state $\cket{\psi(0)}= \cket{\cdots 2020\cdots}$ by using the TEBD algorithm. We found the nonergodic behavior in the time evolution of  the EE and the particle imbalance between the even and odd sites when $U/J$ is large. To better understand this behavior, we compared these results to the real-time dynamics of the initial condition $\cket{\psi(0)}=\cket{\cdots 1010\cdots}$. In this case, we showed that the system approaches thermal equilibrium states. This difference can be understood in terms of the effective spin-1/2 {\it XXZ} models. In the case that the initial state is $\cket{\cdots 2020\cdots}$, the strength of the trapping potential is effectively enhanced when the onsite interaction $U$ increases. Due to this effect, even a small trapping potential can significantly enhance the nonergodic behavior. On the other hand, the strength of the trapping potential is not modified by the interaction in the case that the initial state is $\cket{\cdots 1010\cdots}$. We also calculate the real-time dynamics of the effective Hamiltonian to understand the behavior for a longer time. We showed that the effective spin-1/2 {\it XXZ} model exhibits the nonergodicity on the $O(10^4\hbar/J)$ timescale. We also found the logarithmic growth of the EE in time for $U/J\sim 30$ and saturation of the EE for $U/J\gtrsim 50$. We discussed the relation between this behavior and the dynamics of the local density.

\begin{acknowledgments}
We thank S. Sugawa, T. Tomita, Y. Takahashi, and Y. Takasu for useful discussions. This work was supported by MEXT Quantum Leap Flagship Program (MEXT Q-LEAP) Grant No. JPMXS0118069021 (M.K. and I.D.), JST CREST Grant No. JPMJCR1673 (I.D.), JST FOREST Grant No. JPMJFR202T (I.D.), and JSPS KAKENHI Grants No. JP20K14389 (M.K.), No. JP21H01014 (I.D.), and No. JP18H05228 (I.D.).

\end{acknowledgments}

\appendix
\section{Schrieffer-Wolff transformation}\label{sec:Schrieffer-Wolff transformation}
In this appendix, we discuss the details of the derivation of the effective Hamiltonian by using the Schrieffer-Wolff transformation \cite{Cohen-Tannoudji_book,Bravyi2011}. To perform this, we define the projection operator $\hat{P}$ onto the subspace $\mathcal{H}_{02}$ or $\mathcal{H}_{01}$. We decompose Hamiltonian (\ref{eq:Hamiltonian}) into $\hat{H}=\hat{H}_0+\lambda\hat{V}$, where $\hat{H}_0\equiv \sum_i[V_i\hat{n}_i+(U/2)\hat{n}_i(\hat{n}_i-1)]$ is a nonperturbed part, $\lambda\hat{V}\equiv -J\sum_{\langle i,j\rangle}\hat{a}^{\dagger}_i\hat{a}_j$ is a perturbed part, and $\lambda$ represents a formal perturbation parameter, which we will set $\lambda=1$ in the end of the calculation. The Schrieffer-Wolff transformation transformation is defined by
\begin{align}
\hat{H}'\equiv e^{\hat{S}}\hat{H}e^{-\hat{S}},\label{eq:definition_of_Schieffer_Wolff_transformation}
\end{align}
where $\hat{S}$ is an anti-Hermitian operator. We choose $\hat{S}$ such that $\hat{H}'$ satisfies $\hat{H}'=\hat{P}\hat{H}'\hat{P}+\hat{Q}\hat{H}'\hat{Q}$, $\hat{P}\hat{S}\hat{P}=0$, and $\hat{Q}\hat{S}\hat{Q}=0$, where $\hat{Q}\equiv \hat{1}-\hat{P}$ and $\hat{1}$ is an identity operator. This means that there are no off-diagonal matrix elements between the subspace $\mathcal{H}_{02}$ (or $\mathcal{H}_{01}$) and its complement. The effective Hamiltonian in the subspace $\mathcal{H}_{02}$ or $\mathcal{H}_{01}$ is given by $\hat{H}_{\rm eff}\equiv \hat{P}\hat{H}'\hat{P}$.

We determine the operator $\hat{S}$ by using the second-order perturbation theory. We expand $\hat{S}$ as $\hat{S}=\sum_n\lambda^n\hat{S}_n$. To simplify the notation, we introduce $\hat{X}_{\rm D}\equiv \hat{P}\hat{X}\hat{P}+\hat{Q}\hat{X}\hat{Q}$ and $\hat{X}_{\rm OD}\equiv \hat{P}\hat{X}\hat{Q}+\hat{Q}\hat{X}\hat{P}$, where $\hat{X}$ is an operator. $\hat{S}_n$ is determined by iterative calculations. Within the second order, $\hat{H}'$ is given by
\begin{align}
\hat{H}'&=\hat{H}_0+\lambda\left\{\hat{V}+[\hat{S}_1,\hat{H}_0]\right\}\notag \\
&+\lambda^2\left\{[\hat{S}_2, \hat{H}_0]+[\hat{S}_1, \hat{V}]+\frac{1}{2}[\hat{S}_1, [\hat{S}_1, \hat{H}_0]]\right\}+O(\lambda^3),\label{eq:second_order_H_prime}
\end{align}
Here, we choose $\hat{S}_1$ and $\hat{S}_2$ to cancel the off-diagonal elements of $\hat{H}'$,
\begin{align}
[\hat{S}_1, \hat{H}_0]&=-\hat{V}_{\rm OD},\quad [\hat{S}_2,\hat{H}_0]=-[\hat{S}_1, \hat{V}_{\rm D}].\label{eq:choice_of_S1_S2}
\end{align}
Using these results, we obtain the expression of the effective Hamiltonian,
\begin{align}
\hat{H}_{\rm eff}&=\hat{P}\hat{H}_0\hat{P}+\lambda\hat{P}\hat{V}\hat{P}+\frac{\lambda^2}{2}\hat{P}[\hat{S}_1,\hat{V}]\hat{P}.\label{eq:expression_Heff_second_order}
\end{align}

Here, we consider the case of $\mathcal{H}_{02}$. The zeroth-order effective Hamiltonian becomes
\begin{align}
\hat{P}\hat{H}_0\hat{P}&=\frac{1}{2}MU+\sum_{i=1}^MV_i\hat{P}\hat{n}_i\hat{P}\notag \\
&=\frac{1}{2}MU+\sum_{i=1}^MV_i(2\hat{S}_i^z+1),\label{eq:zeroth_order_H02}
\end{align}
where we used $\cket{2_i}\bra{2_i}=\hat{S}_i^z+1/2$. In this case, we can easily find $\hat{V}_{\rm D}=0$. This means that the first order of the effective Hamiltonian $\hat{P}\hat{V}\hat{P}=0$. From this fact, Eq.~(\ref{eq:choice_of_S1_S2}) becomes $[\hat{S}_1,\hat{H}_0]=-\hat{V}$. Because $\hat{V}$ acts only on nearest-neighbor sites, it is sufficient to consider the matrix elements of the two-site states $\cket{n_i,n_{i+1}}$. The nonzero matrix elements of $\hat{S}_1$ are given by
\begin{align}
\bra{1_i1_{i+1}}\hat{S}_1\cket{0_i2_{i+1}}&=-\bra{0_i2_{i+1}}\hat{S}_1\cket{1_i1_{i+1}}\notag \\
&=\sqrt{2}J_{i,i+1}^+/J,\label{eq:matrix_element_S1_1102}\\
\bra{1_i1_{i+1}}\hat{S}_1\cket{2_i0_{i+1}}&=-\bra{2_i0_{i+1}}\hat{S}_1\cket{1_i1_{i+1}}\notag \\
&=\sqrt{2}J_{i,i+1}^-/J,\label{eq:matrix_element_S1_1120}\\
\bra{1_i3_{i+1}}\hat{S}_1\cket{2_i2_{i+1}}&=-\bra{2_i2_{i+1}}\hat{S}_1\cket{1_i3_{i+1}}\notag \\
&=-\sqrt{6}J_{i,i+1}^+/J,\label{eq:matrix_element_S1_1322}\\
\bra{3_i1_{i+1}}\hat{S}_1\cket{2_i2_{i+1}}&=-\bra{2_i2_{i+1}}\hat{S}_1\cket{3_i1_{i+1}}\notag \\
&=-\sqrt{6}J_{i,i+1}^-/J,\label{eq:matrix_element_S1_3122}
\end{align}
where we used $\hat{S}_1^{\dagger}=-\hat{S}_1$. Using these results, we can obtain the second-order effective Hamiltonian as
\begin{align}
&\frac{1}{2}\hat{P}[\hat{S}_1,\hat{V}]\hat{P}\notag \\
&=\sum_{i=1}^{M-1}\left\{2J^+_{i,i+1}\cket{0_i2_{i+1}}\bra{0_i2_{i+1}}\right.\notag \\
&\left.\quad +2J^-_{i,i+1}\cket{2_i0_{i+1}}\bra{2_i0_{i+1}}\right.\notag \\
&\left.\quad +\tilde{J}_{i,i+1}(\cket{2_i0_{i+1}}\bra{0_i2_{i+1}}+\cket{0_i2_{i+1}}\bra{2_i0_{i+1}})\right.\notag \\
&\left.\quad -6\tilde{J}_{i,i+1}\cket{2_i2_{i+1}}\bra{2_i2_{i+1}}\right\}.\label{eq:second_order_Heff_in_H02}
\end{align}
Finally, using the relations,
\begin{align}
2(\hat{S}_i^x\hat{S}_{i+1}^x+\hat{S}_i^y\hat{S}_{i+1}^y)&=\cket{2_i0_{i+1}}\bra{0_i2_{i+1}}\notag \\
&\hspace{0.1em}+\cket{0_i2_{i+1}}\bra{2_i0_{i+1}},\label{eq:relation_SxSx+SySy}\\
\left(\frac{1}{2}+\hat{S}_i^z\right)\left(\frac{1}{2}+\hat{S}_{i+1}^z\right)&=\cket{2_i2_{i+1}}\bra{2_i2_{i+1}},\label{eq:up-up_term_perturbation_trap}\\
\left(\frac{1}{2}+\hat{S}_i^z\right)\left(\frac{1}{2}-\hat{S}_{i+1}^z\right)&=\cket{2_i0_{i+1}}\bra{2_i0_{i+1}},\label{eq:up-down_term_perturbation_trap}\\
\left(\frac{1}{2}-\hat{S}_i^z\right)\left(\frac{1}{2}+\hat{S}_{i+1}^z\right)&=\cket{0_i2_{i+1}}\bra{0_i2_{i+1}},\label{eq:down_up_term_perturbation_trap}\\
\left(\frac{1}{2}-\hat{S}_i^z\right)\left(\frac{1}{2}-\hat{S}_{i+1}^z\right)&=\cket{0_i0_{i+1}}\bra{0_i0_{i+1}},\label{eq:down_down_term_perturbation_trap}
\end{align}
we obtain the effective Hamiltonian (\ref{eq:second_order_effective_Hamiotonian_for_trap}). 

Next, we consider the case $\mathcal{H}_{01}$. In this case, the zeroth-order effective Hamiltonian becomes
\begin{align}
\hat{P}\hat{H}_0\hat{P}&=\sum_{i=1}^MV_i\hat{P}\hat{n}_i\hat{P}=\sum_{i=1}^MV_i\left(\hat{\tau}_i^z+\frac{1}{2}\right).\label{eq:zeroth-order_H01}
\end{align}
In contract to the $\mathcal{H}_{02}$ case, the first order effective Hamiltonian is nonzero,
\begin{align}
\hat{P}\hat{V}\hat{P}&=-J\sum_{i=1}^{M-1}\hat{P}(\hat{a}_{i+1}^{\dagger}\hat{a}_i+\hat{a}_i^{\dagger}\hat{a}_{i+1})\hat{P}\notag \\
&=-2J\sum_{i=1}^{M-1}(\hat{\tau}_{i+1}^x\hat{\tau}_i^x+\hat{\tau}_{i+1}^y\hat{\tau}_i^y).\label{eq:first_order_term_H01}
\end{align}
In a similar manner, the nonzero matrix elements of $\hat{S}_1$ become
\begin{align}
\bra{0_i2_{i+1}}\hat{S}_1\cket{1_i1_{i+1}}&=-\bra{1_i1_{i+1}}\hat{S}_1\cket{0_i2_{i+1}}\notag \\
&=-\sqrt{2}J^+_{i,i+1}/J,\label{eq:matrix_element_S1_0211_H01}\\
\bra{2_i0_{i+1}}\hat{S}_1\cket{1_i1_{i+1}}&=-\bra{1_i1_{i+1}}\hat{S}_1\cket{2_i0_{i+1}}\notag \\
&=-\sqrt{2}J^-_{i,i+1}/J.\label{eq:matrix_element_S1_2011_H01}
\end{align}
The second-order effective Hamiltonian becomes
\begin{align}
\frac{1}{2}\hat{P}[\hat{S}_1, \hat{V}]\hat{P}&=-2\sum_{i=1}^{M-1}\tilde{J}_{i,i+1}\cket{1_i1_{i+1}}\bra{1_i1_{i+1}}.\label{eq:second_order_effective_Hamiltonian_H01}
\end{align}
Using the above results, we can obtain the effective Hamiltonian (\ref{eq:definition_of_E0_in_H01}).


\end{document}